\documentclass[12pt,preprint]{aastex}
\usepackage{natbib}

\newcommand{\expnt}[2]{\ensuremath{#1 \times 10^{#2}}}   
\newcommand{\gsim}{\gtrsim}
\newcommand{\lsim}{\lesssim}
\shorttitle{Search for Near-IR Counterpart to Cas~A XPS}
\shortauthors{Kaplan, Kulkarni, \& Murray}

\newcommand{\Src}{CXO~J232327.9$+$584842}
\newcommand{\src}{XPS}

\begin{document}
\shorttitle{Search for a Near-IR Counterpart to CXO~J232327.9$+$584842}
\shortauthors{Kaplan, Kulkarni, \& Murray}
\title{Search for an Near-IR Counterpart to the Cas~A X-ray Point Source}
\author{D. L. Kaplan, S. R. Kulkarni}
\affil{Department of Astronomy, 105-24 California Institute of
Technology, Pasadena, California 91125, USA}
\email{dlk@astro.caltech.edu, srk@astro.caltech.edu}
\and \author{S. S. Murray}
\affil{Harvard-Smithsonian Center for Astrophysics, MS-4, 60 Garden
Street, Cambridge, Massachusetts 02138, USA} 
\email{ssm@head-cfa.harvard.edu}

\begin{abstract}
We report deep near-infrared and optical observations of the X-ray point
source in the Cassiopeia~A supernova remnant, \Src.  We have identified a
$J=21.4 \pm 0.3$~mag and $K_{s}=20.5 \pm 0.3$~mag source within the 1-$\sigma$
error circle, but we believe this source is a foreground Pop~II star with
$T_{\rm eff}=2600$--2800~K at a distance of $\approx 2$~kpc, which could
not be the X-ray point source.  We do not
detect any 
sources in this direction 
at the distance of Cas~A, and therefore place 3-$\sigma$ limits of
$R \gsim 25$~mag, ${\rm F675W} \gsim 27.3$~mag, $J\gsim 22.5$~mag and $K_{s}
\gsim 21.2$~mag (and roughly $H \gsim 
20$~mag) on emission from the
X-ray point 
source, corresponding to $M_{R} \gsim 8.2$~mag, $M_{\rm F675W} \gsim
10.7$~mag, 
$M_{J} \gsim 8.5$~mag, $M_{H} \gsim 6.5$~mag, and $M_{K_{s}} \gsim 8.0$~mag,
assuming a distance of 3.4~kpc and an extinction $A_{V}=5$~mag.  
\end{abstract}

\keywords{infrared: stars --- stars: late-type --- supernova: individual (Cassiopeia A)}
\section{Introduction}

Cassiopeia~A (Cas~A) is the youngest Galactic supernova remnant (SNR)
with an age of $\sim 320$~yr, as according to \citet{a80} it is
associated with the explosion observed by \citet{f25} in 1680.
\citet{hrbs00} have found that the elemental abundances in Cas~A are
consistent with those expected from the remnant of a  massive star,
possibly a Wolf-Rayet star \citep{fbb87},
and therefore Cas~A is considered to have been a Type~II supernova.
One therefore expects a compact central remnant, such a neutron
star or black hole, based on the initial mass function of Type~II supernovae
\citep[e.g.][]{ddv98}.
From the first-light images of the {\it Chandra X-ray Observatory}
(CXO), \citet{t99} reported detection of a compact source located
at the apparent center of Cas~A.
The detection of this source, \Src\ (hereafter the X-ray point source
or \src), was later
confirmed in archival {\it 
ROSAT} \citep{ab99} and {\it Einstein} \citep{pz99} data.

The \src\ is located within $5\arcsec$ of the expansion center of
Cas~A \citep{vdbk83}, and given the space density of AGN the chance of
finding one within this distance of the center is quite small.
We convert the count rates from \citet{cph+01} to the 
0.5--2.4~keV band, and get an absorbed flux of
$\approx\expnt{4}{-13}\mbox{ ergs s}^{-1}\mbox{ cm}^{-2}$.
Comparing this with the AGN $\log N$-$\log S$ relation from
\citet{gss+96}, we would expect $\sim 0.4\mbox{ AGN
deg}^{-2}$, or $\sim \expnt{2}{-6}$~AGN of this flux at the center of
Cas~A.  It is thus extremely improbable that the \src\ is an AGN, a
fact further confirmed by its relatively steep spectrum \citep{cph+01}.

Therefore, it is generally believed that the \src\
is associated with the remnant of the Cas~A progenitor
\citep{cph+01}.  
The X-ray spectrum of the \src, as determined by \citet{pza+00} and
\citet{cph+01}, can be fitted by a power-law with a photon index $\sim 3$.
Other acceptable fits include thermal bremsstrahlung
($kT^{\infty}\approx 1.7$~keV), blackbody ($kT^{\infty} \approx
0.5$~keV, $R^{\infty}\approx 0.5$~km), or neutron star atmospheres 
($kT^{\infty}\approx 0.4$~keV, $R^{\infty}\approx 0.8$~km for the
model of 
\citealt{hh98}; $kT^{\infty}\approx 0.27$~keV, $R^{\infty}\approx
2$~km for the model of \citealt{zps96}).  

The nature of the \src\ is unclear.  However, we have an idea as to
what it is not.  The spectral index of the \src\ is significantly
steeper than those typical for young X-ray pulsars, its 
luminosity is $\gsim 10^{2}$ times less than those of young X-ray
pulsars,  and there is no evidence for a synchrotron nebula
\citep{mcd+00}.  The spectrum is similar to that of an anomalous X-ray
pulsar \citep[AXP; see][]{m99}, but the X-ray luminosity is 
 at least several (if not 10--100) times fainter than that
typical for AXPs.  The \src\ is cooler but much 
more luminous than isolated neutron stars \citep{m00}.

Furthermore, there have not been any detections of optical
\citep{vdbp86,rws01} or radio \citep[][ and references therein]{mcd+00}
emission from the \src, nor have  X-ray pulsations been
detected \citep{cph+01}, though the current limits are not very
constraining.  Therefore, the \src\ is almost certainly not a 
young pulsar similar to the Crab.  Theories as to its identity range
from a cooling neutron star emitting from polar caps to an accreting
black hole \citep{untm00,pza+00}. 

From measures of line ratios in the Cas~A remnant, \citet{s71} finds
the extinction to be $A_{V}=4.3$~mag
in the direction of one filament.  Later radio studies found significant
variations of $A_{V}$ on scales of $\sim 1\arcmin$, and overall values ranging
from 4--5~mag for the north and northeastern rim and $\gsim 5$--6~mag
for the rest of the SNR \citep{tch85}.  Similarly, \citet{hf96} find
extinction values of 4.6--5.4~mag across the northern portion
(assuming $R_{V}=3.1$).  We will therefore adopt a middle value of
$A_{V}\approx5$~mag.   We assume that Cas~A and the \src\ are at a distance of 
$3.4^{+0.3}_{-0.1}$~kpc \citep{rhfw95}, which we parameterize as
$D=3.4d_{3.4}$~kpc. 

In this letter we report on optical/near-IR searches for a counterpart
to the \src.  We believe that given the unknown nature of \src,
searches at all wavelengths are warranted and even upper limits may
constrain the nature of this enigmatic source.  The paper is organized as follows: in
Section~\ref{sec_obs} we detail our observations and reduction
techniques.  Section~\ref{sec_results} contains a description of the
results, while Section~\ref{sec_analysis} presents an analysis of these
results.  Finally, a discussion and conclusions are in Section~\ref{sec_discuss}.

\section{Observations}
\label{sec_obs}
\subsection{Cas-A Central Point Source Position}
The SNR Cas A was observed several times with the CXO.
After the initial detection in the first-light images \citep{t99}, a long HRC-I observation
was obtained on 1999 December 20,  and a  third observation with the
HRC-S in imaging mode  was taken on 2000 October 5. A discussion of the
results from this observation is in preparation (Murray et al. 2001);
here we provide only the source location information. Table~\ref{tab:xray}
gives the point source locations and estimated uncertainties (including
estimates of systematic errors). We estimate that the overall positional uncertainty
for all of these observations is $1\farcs0$ (1-$\sigma$).

\subsection{Optical and Near-IR Observations}
The observations were carried out primarily with the Near Infrared
Camera \citep[NIRC;][]{ms94} mounted on the 10-m Keck~I telescope,
augmented with archival HST/WFPC2 images.  We
also took 
auxiliary optical and infrared calibration images with the COSMIC
imager on the Palomar 5-m  telescope (P200), the PFIRCAM
\citep{jbvb+94} infrared
imager on the P200, and the P60CCD optical imager on the Palomar 1.5-m
telescope (P60).  A summary of the instruments, filters, exposures,
and conditions is listed in Table~\ref{tab_obs}.

The optical data were reduced with the standard {\tt IRAF} {\tt
ccdred} package.  The images were bias subtracted, flat-fielded,
registered, and
co-added.  The infrared data were reduced with custom {\tt IRAF}
software.  The images were dark subtracted, flat-fielded, and corrected
for bad pixels and cosmic rays.  We then made object masks, which were
used in a second round of flat-fielding to remove holes from the
flats.  The data were then registered, shifted, and co-added.  The HST
images were processed using the standard drizzling procedure \citep{fm98}.

The data from the P60CCD were used as the astrometric reference.  We
matched 36 non-saturated stars to those from the USNO-A2.0 catalog
\citep{m98}. 
Using the task {\tt ccmap} we computed a transformation solution,
giving $0\farcs2$ residuals (all astrometric residuals are 1-$\sigma$
for each coordinate unless otherwise indicated).  We then used this 
solution to fit stars on the COSMIC images.  Using 37 stars, we
again obtained $0\farcs2$ residuals.

We then used 15 stars on the COSMIC images to fit the HST image,
getting $0\farcs07$ residuals.  This solution was then used for the
infrared images, fitting 10 stars on the NIRC images with $0\farcs05$ residuals.
This gives $0\farcs4$ position uncertainties relative to
the ICRS, assuming the uncertainties intrinsic to the USNO-A2.0 are
$0\farcs3$ \citep[for each axis;][]{m98}
We then transfered this solution to the PFIRCAM images
($0\farcs3$ residuals), but as this is only a photometric reference
the absolute position is not important.

For the optical photometry, we used $V$, $R$, and $I$ observations of
the standard
fields\footnote{\url{http://cadcwww.dao.nrc.ca/cadcbin/wdb/astrocat/stetson/query/}} 
Landolt~110, NGC~7790, and PG~1657 \citep{l92,s00} carried out 
with the P60CCD.  We fit the observations over the whole night using
airmass corrections and first-order color terms, and measured the $R$
zero-point magnitude.  We then examined 25 stars
on the Cas~A images common to both the P60CCD and COSMIC images, and
from this determined the zero-point for the photometric night.  From
these data we also determined the limiting 
magnitude to be $R\sim 25$~mag.

For the infrared photometry, we used 3 observations of the faint UKIRT
standard stars FS~29 and FS~31  \citep{ch92} taken with the PFIRCAM.  These
observations were used to determine $J$, $H$, and $K_{s}$ zero-points
(we assumed the $K_{s}$ magnitudes were the same as the $K$
magnitudes, as the correction is typically $\lsim 0.01$~mag: much
smaller than our uncertainties; \citealt{pmk+98}).  From these images
we then found 5 stars common to the PFIRCAM and NIRC  images, and
determined zero-point magnitudes for NIRC.

\section{Results}
\label{sec_results}
\begin{figure*}
\epsscale{1.0}
\centerline{
\vbox{
\hbox{\plottwo{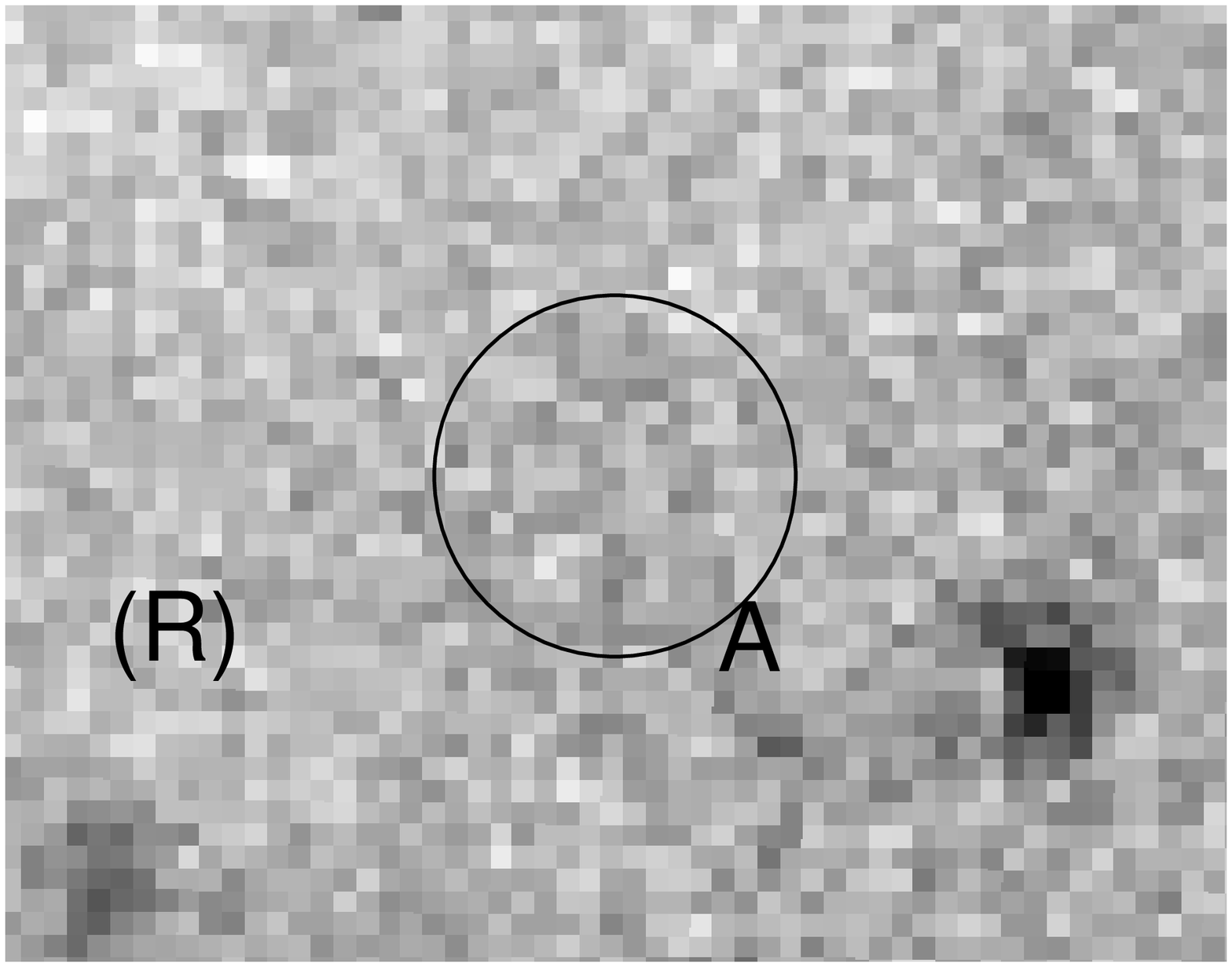}{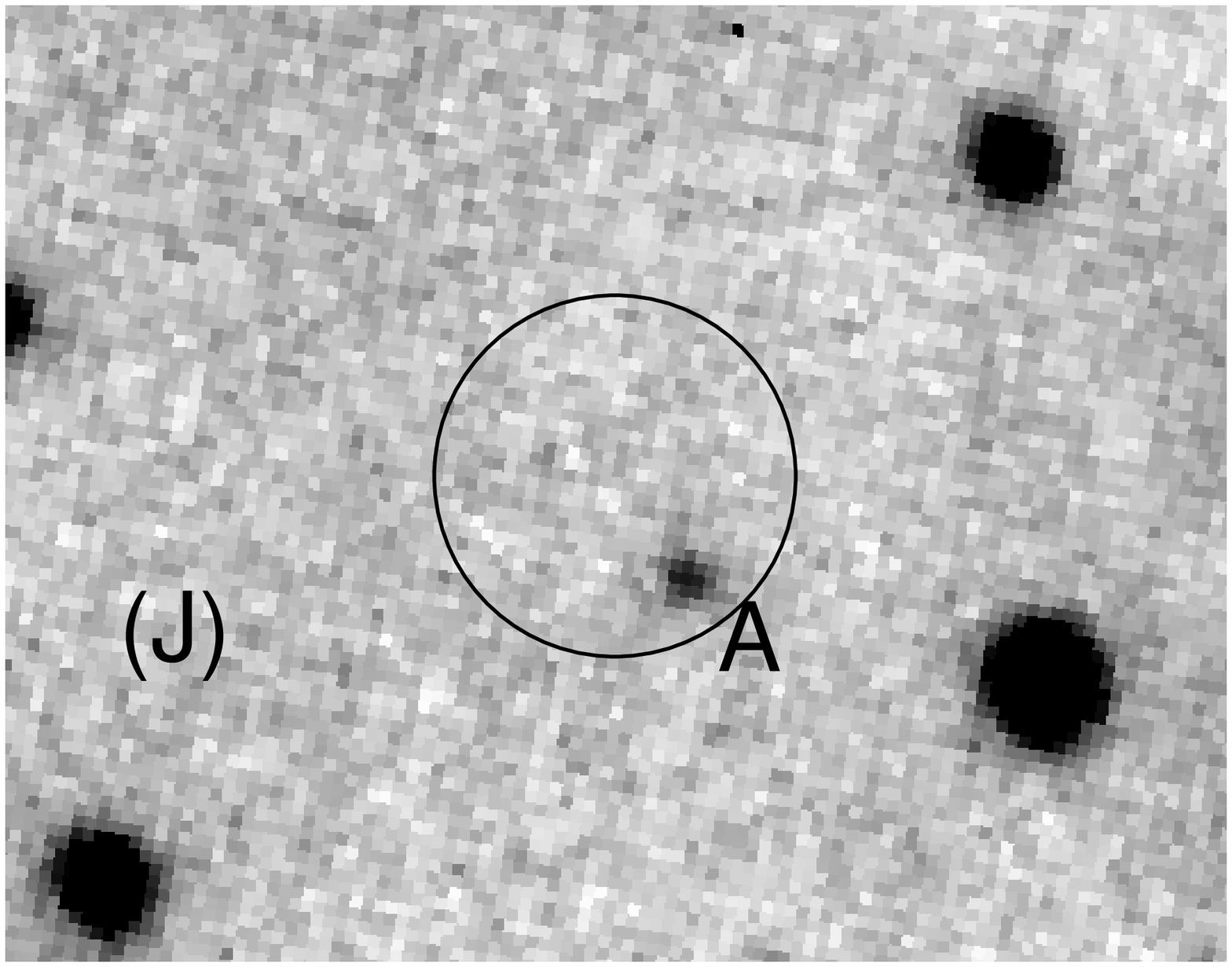}}
\hbox{\plottwo{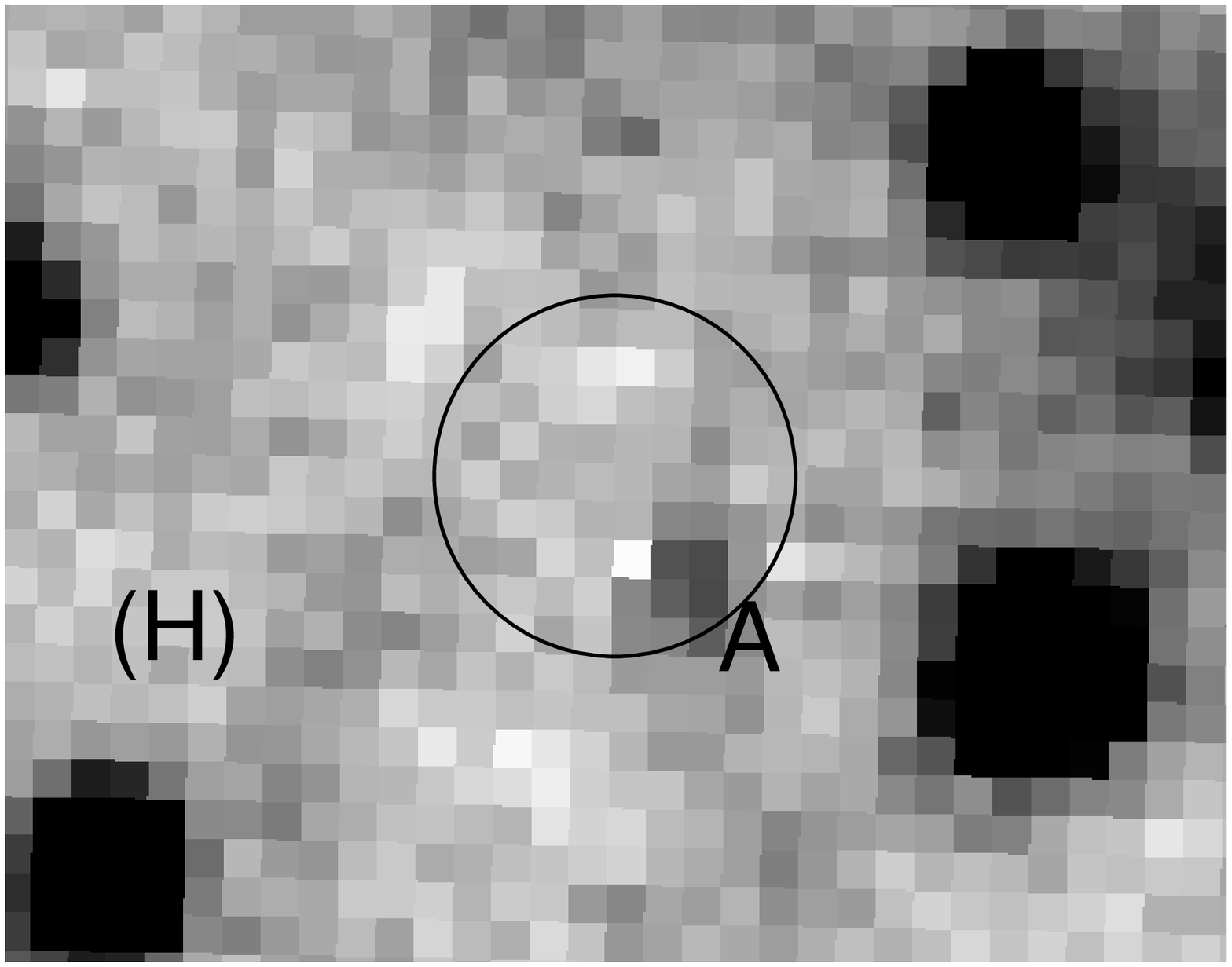}{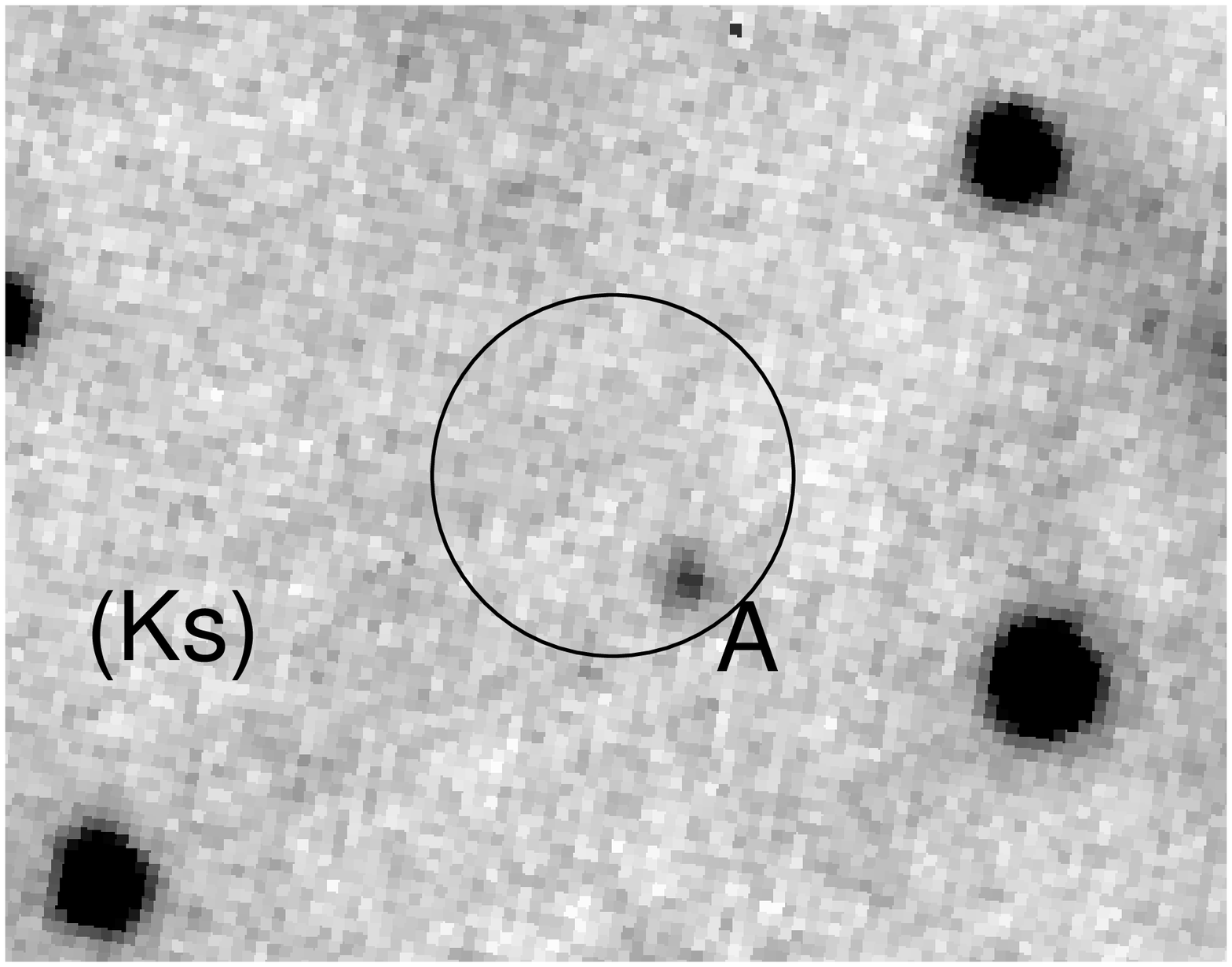}}}}
\caption{Images of the region around the \src.  They are: $R$-band
(COSMIC; 
upper left); $J$-band (NIRC; upper right); $H$-band (PFIRCAM; lower left);
$K_{s}$-band (NIRC; lower right).  A $2\farcs3$ radius circle (90\%
confidence) is drawn around
the position of the \src, and candidate source A is indicated.  North
is up, and East is to the left.  The images are $\approx 15\arcsec$ on
each side.}
\label{fig_images}
\end{figure*}

We searched for a counterpart to the X-ray point source, at position
$\alpha(J2000)=23^{\rm h}23^{\rm m}27\fs857$, $\delta(J2000)=+58\degr
48\arcmin 42\farcs77$, with $1\farcs0$ uncertainty (Table~\ref{tab:xray}).  See
Figure~\ref{fig_images} for the separate optical/IR images.  There was no
source on COSMIC images, giving $R \gsim 25$~mag (3-$\sigma$ limit) for
any possible 
counterpart (this agrees with the previous limit of $R\gsim 24.8$~mag and
$I\gsim 23.5$~mag;
\citealt{vdbp86}).  On the NIRC, PFIRCAM, and HST images there
was a source $1\farcs7$ away from the X-ray position, at
$\alpha(J2000)=23^{\rm h}23^{\rm m}27\fs78$,
$\delta(J2000)=+58\degr48\arcmin41\farcs2$ ($\pm 0\farcs4$ in each
coordinate).  Given the  
astrometric uncertainties, the overall  position uncertainty is
$1\farcs1$ in each axis, so this source is 1.5-$\sigma$ away from the nominal
position.  We label this source A, and consider it as a
potential candidate counterpart or companion to the X-ray source.  The 
magnitudes of source A are ${\rm F675W}=26.7 \pm 0.2$~mag,
$J=21.4 \pm 0.3$~mag, 
$H \approx 20.5 \pm 0.8$~mag, and $K_{s}=20.5 \pm 0.3$~mag.
There are no other
sources within the $2\farcs3$ radius 90\% confidence circle.

\section{Analysis}
\label{sec_analysis}
\begin{figure*}
\epsscale{1.0}
\centerline{
\hbox{\plottwo{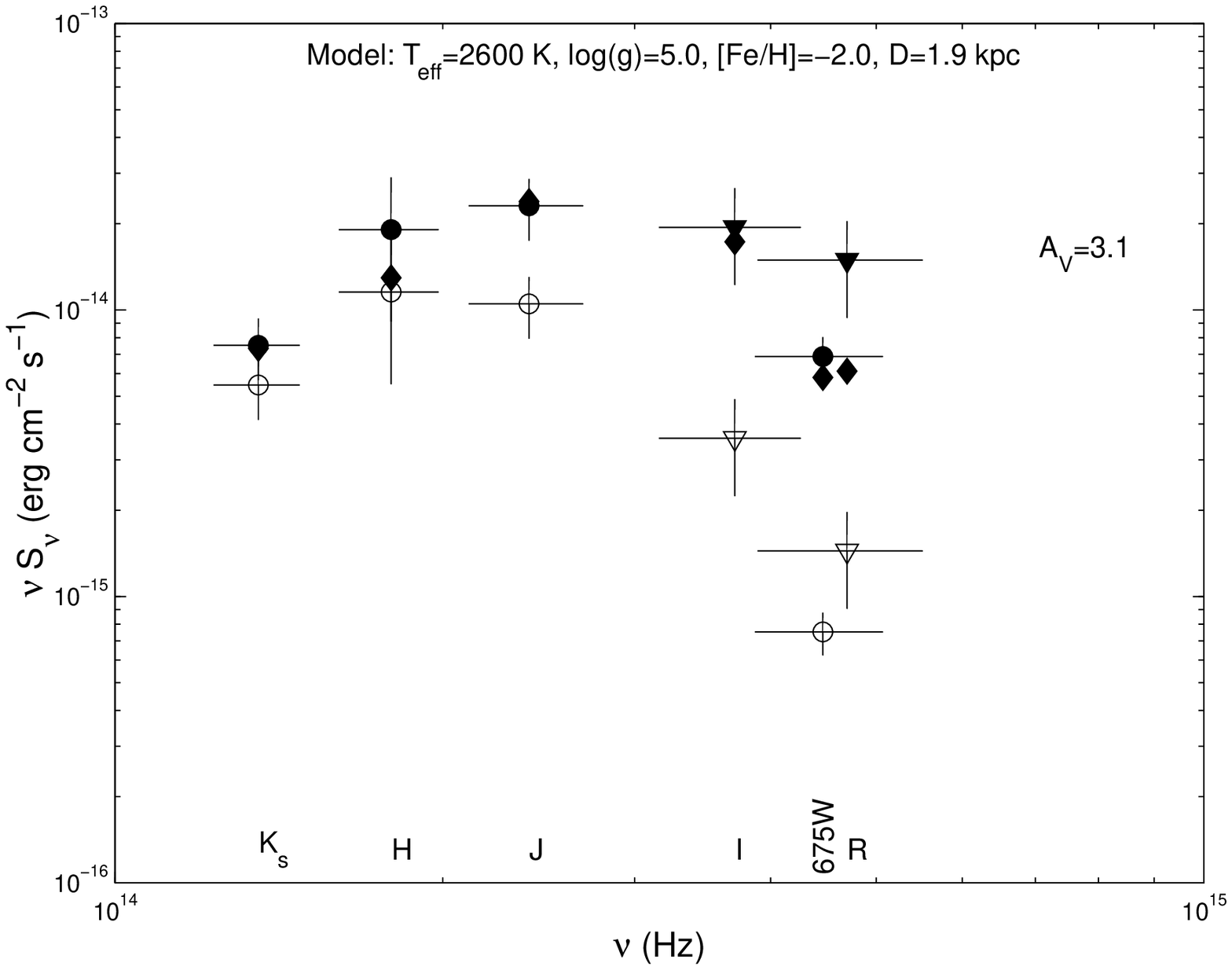}{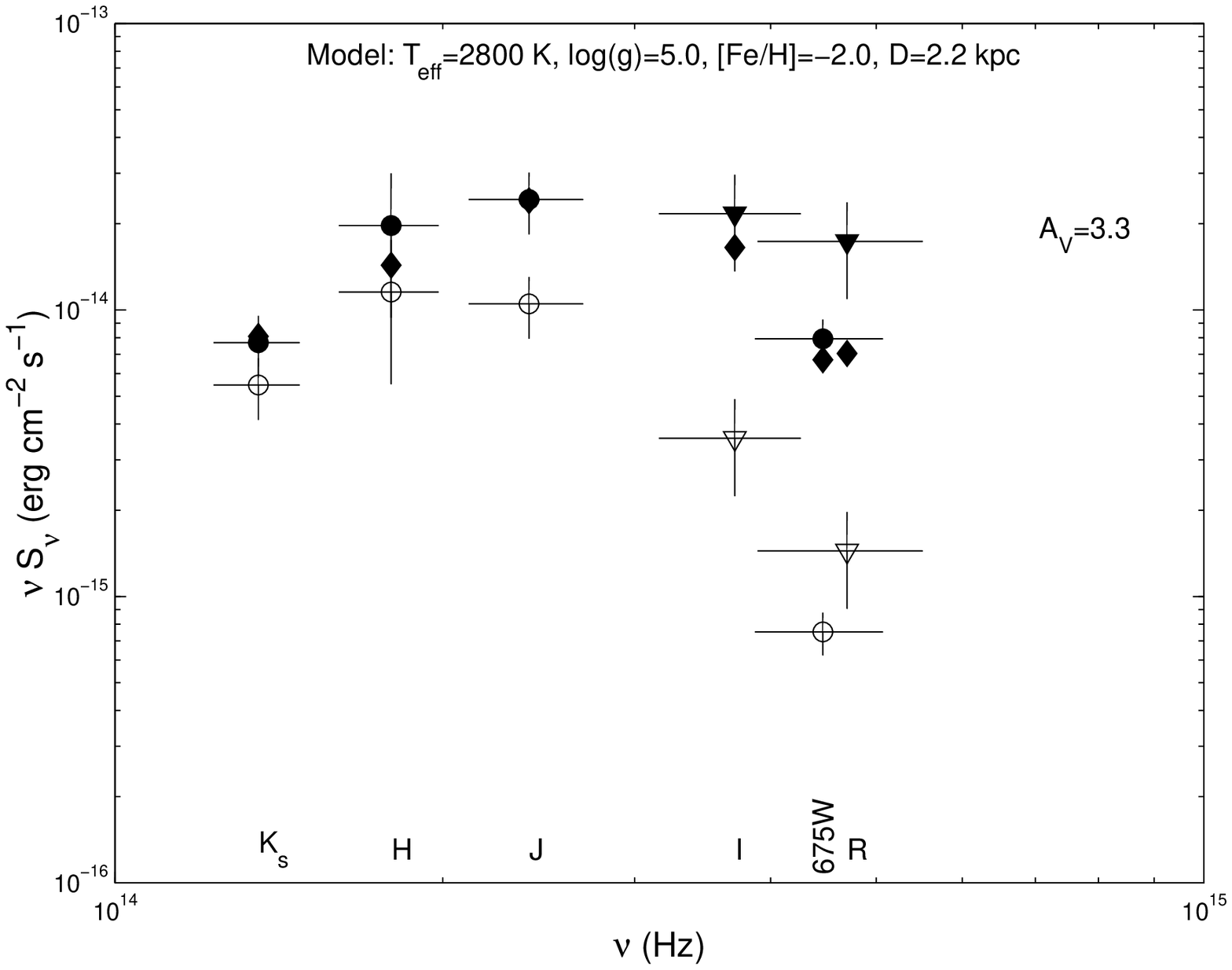}}}
\caption{Spectral energy distribution for source~A, a foreground star from the Cas~A X-ray
error circle, and best-fit data for two different stellar models:
$T_{\rm eff}=2600$~K (left); $T_{\rm eff}=2800$~K 
(right).  The
open symbols are the observed data, the 
filled circles and limits those corrected for reddening, and the
diamonds the model values.  The model parameters, from \citet{bcp98},
are listed on the figures.}
\label{fig_sed}
\end{figure*}

Using the reddening  and
zero-point calibration data from \citet{bcp98}, we plot the
spectral energy distribution (SED) for source A  in
Figure~\ref{fig_sed}.  This incorporates both the  detections
and   limits.

To determine if source A could be a star, we compared model stellar colors
from \citet{bcp98} with our data.  We fitted for
three parameters: the visual extinction $A_{V}$, the distance in kpc
$D_{\rm kpc}$, and
the stellar model (which includes the effective temperature $T_{\rm
eff}$, the surface gravity $g$, and the metallicity [Fe/H]).  We
assumed that the star would be a zero-age main-sequence star such that
$\log(R/R_{\sun})=0.7 \log(M/M_{\sun}) -0.1$ \citep{hh81}, and used
the bolometric corrections and reddening from  
\citet{bcp98} to find the expected magnitudes.   To
account for the upper limits in our fitting, we minimized a modified
$\chi^{2}$ statistic, such that
\begin{equation}
\chi^{2} = \sum_{i}^{\rm Detect} \left(\frac{m_{i}-
m_{i,{\rm mod}}}{\sigma_{i}}\right)^{2} + \sum_{i}^{\rm non-Detect}
\cases{0 &if $m_{i,{\rm mod}} \geq m_{i}$; \cr
\left(\frac{m_{i,{\rm mod}}-m_{i}}{\sigma_{i}}\right)^{2}
&otherwise. \cr}
\label{eqn_chi}
\end{equation}

Here, $i$ runs over the different filters, $m_{i}$ is the observed
magnitude or limit for that filter, $m_{i,{\rm mod}}$ is the model
magnitude, and $\sigma_{i}$ is the uncertainty.  
The model uses standard Vega-based magnitudes, where the HST data do
not.  Therefore we converted the HST magnitude to the Vega-based
system using $R-\mbox{F675W}=-1.05$~mag, appropriate for sources of
this color\footnote{http://www.stsci.edu/instruments/wfpc2/Wfpc2\_phot/wfpc2\_cookbook.html}.
We do not incorporate
model uncertainties into this statistic.  Minimizing this $\chi^{2}$
seeks the best model that comes close in magnitude to the detections
while remaining fainter than the non-detections.  A full-fledged
Bayesian analysis \citep[e.g.][]{gl92,cc97} would be more accurate,
but we only wish to demonstrate the plausibility of model fits, not
assign specific probabilities to different models.

Given the number of variables, this fit is somewhat unconstrained.
We restrict  the 
extinction and distance to reasonable values ($0.5 \lsim A_{V} \lsim
8$~mag, $0.5 \lsim D_{\rm 
kpc} \lsim 5$), and  fit for $\log(g)=5.0$ (appropriate for late M
stars; \citealt{hh81}).  In addition, we require that $A_{V}$ 
roughly scale with $D$, excluding models that are very distant but
have almost no extinction.  We find that our detections and limits
are entirely 
consistent with a 
cool (M6--8), Pop~II, main-sequence star, which is  between the Earth
and  Cas~A.    
Good fits are obtained for stars with $T_{\rm eff}=2600$--2800~K,
[Fe/H]$=-2.0$, 
$D_{\rm kpc} =1.8$--2.0~kpc, and $A_{V}=3.1$--3.2~mag (see
Figure~\ref{fig_sed} for examples).  We do not give the $\chi^{2}$
value or formal confidence 
regions, as the $\chi^{2}$ in Equation~\ref{eqn_chi} is somewhat
contrived and the models for stars this cool are not well determined
\citep{bcp98}, but Figure~\ref{fig_sed} demonstrates the plausibility
of  the fits.  That there is a star within $1\farcs7$ of the \src\ is
quite believable:  the
theoretical star-count models of \citet{nim+00} give $1.5 \times 10^{6} \mbox{
stars deg}^{-2}$ of the appropriate colors with $J \leq 22.5$~mag,
leading to a false coincidence rate of 1.0.
The best-fit star has $R=0.2 R_{\sun}$, $M=0.1 M_{\sun}$,
and $L=0.004 L_{\sun}$.   Slightly  deeper $I$ band observations should
be able to verify the classification of source A.

As one might expect, there is a significant anti-correlation in the
fits 
between values of $D_{\rm kpc}$ and $A_{V}$, with $\pm
0.25$~kpc and $\pm 0.5$~mag variations giving reasonable
fits, but the fits for the range of likely
extinctions for Cas~A (4--6~mag) at 3.4~kpc are definitely poor.

Assuming that source A is a late-type star, we examine whether it
could be associated with the \src, implying that both are in the
foreground and that the \src\ is not
associated 
with Cas~A.  From
\citet{kc00}, we see that for 
a star with $B-V \geq 1.8$ (from the model for source A), the X-ray
luminosity is $L_{\rm X,star}($0.1--2.4~keV$)\lsim
10^{28} \mbox{ ergs s}^{-1}$ \citep[also][]{jjj+00,mmp00}, giving
unabsorbed (denoted by superscript U) X-ray-to-infrared flux ratios of
$f_{\rm X}^{\rm U}/(\nu_{J}
f^{\rm U}_{\nu, J})=6\times 10^{-4}$ and  $f_{\rm X}^{\rm U}/(\nu_{K_{s}}
f^{\rm U}_{\nu, K_{s}})=2\times 10^{-3}$ for such a star.  Converting the flux of the \src\ 
to the ROSAT passband (using 
W3PIMMS\footnote{\url{http://heasarc.gsfc.nasa.gov/Tools/w3pimms.html}}),
it has ratios of $f_{\rm X}^{\rm U}/(\nu_{J}
f^{\rm U}_{\nu, J})=34$, $f_{\rm X}^{\rm U}/(\nu_{K_{s}}
f^{\rm U}_{\nu, K_{s}})=111$, which are drastically different.
In addition, \citet{pza+00} and \citet{cph+01} did not observe any
variability from the \src, unlike late-type stars that can vary by factors of
$\sim 10^{2}$ on small time scales \citep{mmp00}.
Source~A therefore could not emit the X-rays
observed from the \src.

We conclude that the \src\ was not detected, and add ${\rm F675W} \gsim
27.3$~mag, $J \gsim 22.5$~mag and $K_{s}
\gsim 21.2$~mag (3-$\sigma$), along with a rough limit of $H \gsim 20$~mag, to
the previously mentioned limits.   

\section{Discussion \& Conclusions}
\label{sec_discuss}
\begin{figure*}
\plotone{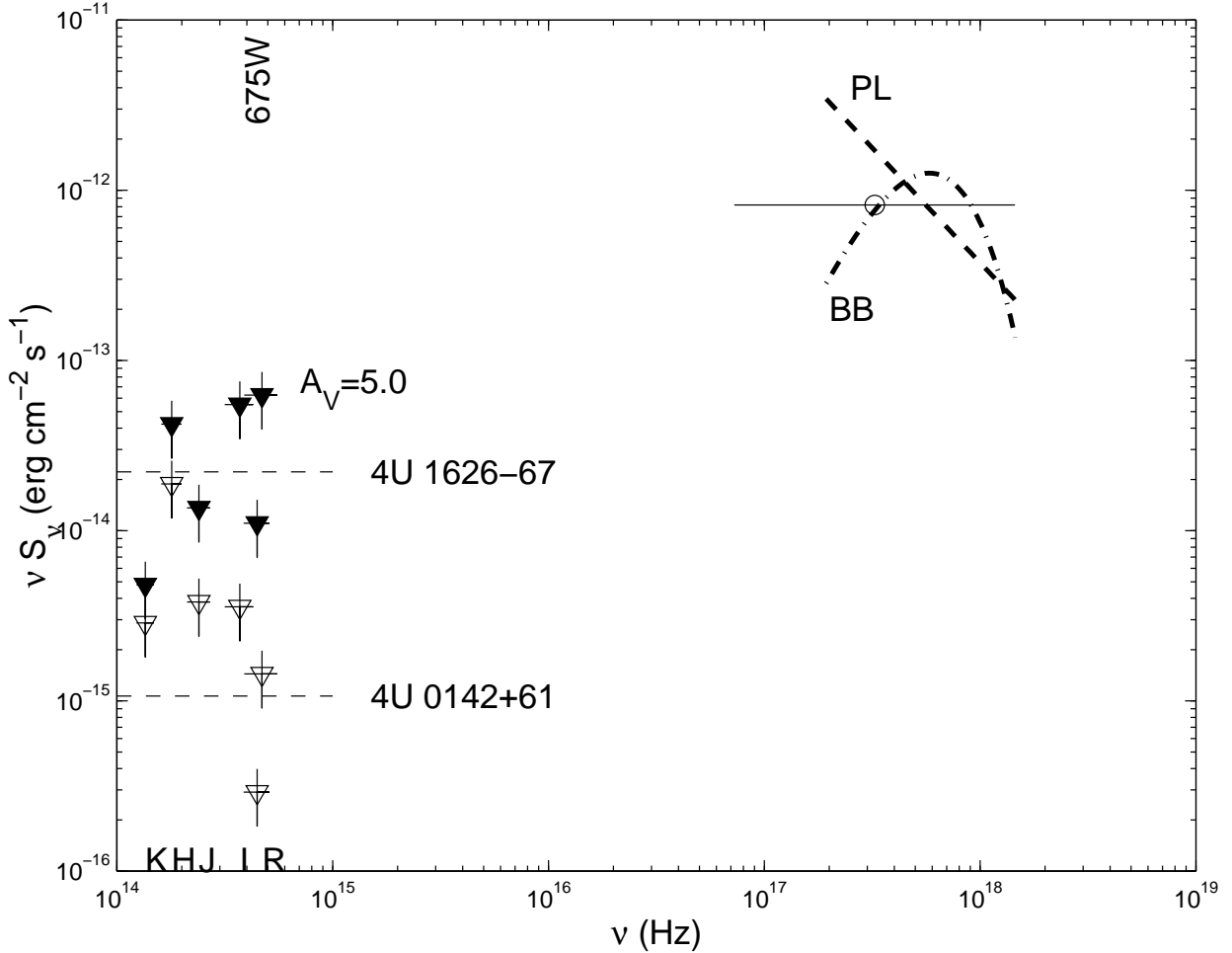}
\caption{Spectral energy distribution for the \src.  This
incorporates optical limits (this work) and X-ray data
\citep{pza+00}.  The open triangles are the measured values, while the
filled triangles are those corrected with $A_{V}=5.0$~mag.  The open
circle is the measured CXO flux, while the thick lines are model
spectra corrected for absorption: power-law (PL; dashed) and blackbody
(BB; dash-dotted).  The thin dashed lines are derived from the
unabsorbed X-ray-to-$R$-band ratios of the AXP 4U~0142$+$61
\citep[][with $A_{V}=4.4$~mag]{hvkk00} and the
very close X-ray binary 4U~1626$-$67 \citep[][with $A_{V}=0.2$~mag]{c98},
assuming a 0.5--10~keV luminosity of 
$10^{34}\mbox{ ergs s}^{-1}$ for the \src.  We do not plot the more complicated 
atmosphere models from \citet{pza+00} or \citet{cph+01}.}
\label{fig_allsed}
\end{figure*}

Based on a synthesis of CXO, {\it ROSAT}, and {\it Einstein} data,
\citet{pza+00} fit the X-ray spectrum of the \src.  The absorbed 
flux is $8.2\times 10^{-13}\mbox{ ergs cm}^{-2}\mbox{
s}^{-1}$ in the 0.3--6.0~keV range
\citep{pza+00}.  Power-law and pure blackbody models give good fits to
the absorption-corrected data,
and are plotted as representative X-ray spectra in
Figure~\ref{fig_allsed}.  These results are similar to those from
\citet{cph+01}.  \citet{pza+00} prefer the results of a
H/He polar-cap model with a cooler Fe surface, but all we wish to
illustrate is that blackbody models are consistent with the optical
limits, while power-law models require a break between the X-ray and
optical bands.

In Figure~\ref{fig_allsed} we also plot the expected optical magnitudes of
representative X-ray sources (an AXP and a tight X-ray binary) for
comparison.  These magnitudes are derived by taking the
X-ray-to-optical flux ratios for these objects and scaling them to the
X-ray 
flux of the \src.  We can likely reject sources like 4U~1626$-$67
\citep{c98} from consideration, but the extrapolation of the AXP
4U~0142$+$61 \citep{hvkk00} 
is consistent with the current limits.

Giving the presumed distance and  reddening, our limits translate to
$M_{R} \gsim 8.2$~mag, $M_{\rm F675W} \gsim 10.7$~mag, $M_{J} \gsim 8.5$~mag, $M_{H} \gsim 6.5$~mag, and
$M_{K_{s}} \gsim 8.0$~mag.   We find the observed X-ray-to-infrared flux
ratios to be $f_{\rm X}/(\nu_{\rm F675W}f_{\nu, \rm
F675W})\gsim 2872$, $f_{\rm X}/(\nu_{J} f_{\nu, J}) \gsim 212$, $f_{\rm X}/(\nu_{K_{s}}
f_{\nu, K_{s}}) \gsim 280$ (the X-ray flux is in the 0.3--6.0~keV band).
If we correct for interstellar absorption, we find unabsorbed
ratios
of $f^{\rm U}_{\rm X}/(\nu_{\rm F675W}f^{\rm U}_{\nu, \rm F675W})
\gsim 231$, $f^{\rm U}_{\rm X}/(\nu_{J}
f^{\rm U}_{\nu, J}) \gsim 166$ and  $f^{\rm U}_{\rm X}/(\nu_{K_{s}} f^{\rm
U}_{\nu, K_{s}}) \gsim 467$, using the X-ray flux from
\citet{cph+01}.  These flux ratios, larger than those inferred
previously, tighten constraints on the identity of the \src\
\citep[e.g.][]{untm00,pza+00}. 

\acknowledgements
We would like to thank T. Nakajima for supplying star-count models,
and J. Cordes for sharing observing time.
DLK is supported by the Fannie and John Hertz Foundation and SRK by 
NSF and NASA.  DLK thanks the ITP at Santa Barbara,
where part of the work presented here was done, for hospitality.  The
ITP is supported by the National Science Foundation under Grant
No. PHY99-07949.  Data presented herein were obtained at the W.M.
Keck Observatory, which is operated as a scientific partnership among
the California Institute of Technology, the University of California
and the National Aeronautics and Space Administration.  The
Observatory was made possible by the generous financial support of the
W.M. Keck Foundation.   Data are also based on observations with the
NASA/ESA Hubble Space Telescope, obtained from the data Archive at the Space Telescope
Science Institute, which is operated by the Association of
Universities for Research in Astronomy, Inc. under NASA contract No. NAS5-26555

\bibliographystyle{apj}



\begin{deluxetable}{c c c c l l}
\tablecaption{Cas A X-ray Observation Summary\label{tab:xray}}
\tablehead{
\colhead{OBSID} & \colhead{Date} & \colhead{CXO}
 & \colhead{Exposure} & \colhead{RA\tablenotemark{a}} & \colhead{Dec\tablenotemark{a}} \\
 & &\colhead{Instrument} & \colhead{(ks)} & \colhead{(J2000)} &
\colhead{(J2000)} \\
}
\startdata
214& 1999 Aug 20& ACIS S3& 6& $23^{\rm h}23^{\rm m}27\fs94$ & $+58\degr48^{\prime}42\farcs4$\\
1505& 1999 Dec 20& HRC-I& 50& $23^{\rm h}23^{\rm m}27\fs88$&  $+58\degr48^{\prime}42\farcs1$\\
1857& 2000 Oct 05& HRC-S& 50& $23^{\rm h}23^{\rm m}27\fs75$&  $+58\degr48^{\prime}43\farcs8$\\
\tableline
{Average}& & & & $23^{\rm h}23^{\rm m}27\fs857$ & $+58\degr48^{\prime}42\farcs77$\\
{Uncertainty}\tablenotemark{b}& & & & $\phm{23^{\rm h}23^{\rm m}2}0\fs097$& $\phm{+58\degr48^{\prime}4}0\farcs91$\\
\enddata
\tablenotetext{a}{The individual source positions were calculated as centroids of the
event distributions taken within a $1\farcs0$ radius circle about
the location and iterated until the centroid location shifted by less
than $0\farcs1$.}
\tablenotetext{a}{Uncertainties are 1-$\sigma$.}
\end{deluxetable}

\begin{deluxetable}{l l l c c l}
\tablecaption{Cas~A Optical/Near-IR Observation Summary\label{tab_obs}}
\tablecolumns{6}
\tablewidth{36pc}
\tablehead{
\colhead{Date} & \colhead{Telescope /} & \colhead{Observer} & \colhead{Band} &
\colhead{Exposure} & \colhead{Conditions} \\
& \colhead{Instrument} & & & \colhead{(s)} & \\}
\startdata
2000~Jan~22 & HST/WFPC2 & R.~Fesen & F675W & 4000 & \nodata \\
2000~Jun~27 & Keck~I/NIRC & S.~Kulkarni & $J$ & 1600 & slight cirrus \\
& & & $K_{s}$ & 2364 & \\
2000~Jul~04 & P200/COSMIC & P.~Mao &  $R$ & 1010 & photometric\\
2000~Jul~05 & P200/COSMIC &  P.~Mao & $R$ & 1000 & high cirrus\\
2000~Jul~24 & P60/P60CCD & D.~Kaplan &  $R$ & 150 & photometric\\
& & & $I$ & 150 & \\
2000~Sep~06 & P200/PFIRCAM &  D.~Kaplan / & $J$ & 1680 & photometric \\
 & & J. Cordes & $H$ & 1680 & \\
 & & & $K_{s}$ & 1120 & \\
\enddata
\end{deluxetable}

\end{document}